\documentstyle{article}
\input pz.sty
\input epsf.sty
\begin{document}

\PZhead{2}{30}{2010}{10 March}

\PZtitletl{SN 2008\lowercase{fv}: the third type I\lowercase{a}}{supernova 
in NGC 3147}

\PZauth{D. Yu. Tsvetkov$^1$, L. Elenin$^2$}
\PZinst{Sternberg Astronomical Institute, University Ave., 13,
119992 Moscow, Russia}
\PZinst{Keldysh Institute of Applied Mathematics RAS, Miusskaya sq., 4, 
125047 Moscow, Russia}

\begin{abstract}
Multiple outbursts of type Ia SNe in one galaxy present
a unique opportunity to study the homogeneity of these
objects.
NGC 3147 is only the second known galaxy with three SNe Ia,
another one is NGC 1316. We present 
CCD $UBVRI$ photometry for SN Ia
2008fv and compare the light and color curves of
this object with those for SNe Ia discovered earlier in NGC 3147: 
1972H and 1997bq.
The photometric properties of SNe 1997bq and 2008fv are nearly
identical, while SN 1972H exhibits faster declining light curve.  
\end{abstract}
\bigskip
\bigskip
\PZsubtitle{SN 2008fv}

SN 2008fv was discovered by Itagaki on unfiltered CCD images exposed 
with a 0.6-m reflector around September 27.78 UT at magnitude 16.5.
The new
object was located at 
$\alpha  = 10\hr16\mm57\sec.28, \delta = +73\deg24\arcm36\arcs.4$
(equinox 2000.0), which is $16\arcs$ east and $34\arcs$ north of the 
center of the galaxy NGC 3147 (Nakano, 2008).  

Challis (2008) reported on behalf of the CfA Supernova Group that a spectrum 
(range 350-880 nm)
of SN 2008fv, obtained on September 30 UT with the MMT at Mt. Hopkins,
shows it to be a 
normal type-Ia supernova about one week before maximum light.
This is the fourth SN discovered in NGC 3147, the former were 
SN I 1972H (Goranskij, 1972; Barbon et al., 1973),  
SN Ia 1997bq (Laurie, Challis, 1997; Jha et al., 2006) and SN Ib 2006gi 
(Itagaki, 2006; Elmhamdi et al., 2010).

\medskip

We started photometric monitoring of SN 2008fv soon  
after discovery, on October 3, 
with the remotely controlled telescope of the Tzec Maun Observatory.
Later we also observed the SN with four other telescopes at three different
locations.   
The data on the telescopes and detectors 
are presented in Table 1.

\begin{table}
\caption{Telescopes and detectors used for observations}\vskip2mm
\begin{tabular}{lcrllll}
\hline
Telescope & D(cm) & F(cm) & CCD camera & Filters & Location & Code \\
\hline
Maksutov- & 35 & 133  & SBIG ST-10E & $VR$ & Mayhill, New  & TM35\\
Newton &  &   & &  & Mexico, USA & \\
Cassegrain      & 70 & 1050 & Apogee AP-7p & $UBVRI$ & Moscow, Russia & M70 \\
Cassegrain      & 60 & 750  & Apogee AP-47p & $BVRI$ & Nauchny, Crimea, & C60 \\
                                          &  &   & &  & Ukraine & \\
Maksutov        & 50 & 200  & Meade Pictor & $VRI$ & Nauchny, Crimea, & C50\\
                &    &      &416XT         &       & Ukraine & \\
Newton          & 50 & 250  & SBIG  & $UBVRI$ & Tatranska Lomnica,  & S50 \\
                &    &      & ST-10XME  &  & Slovakia & \\

\hline
\end{tabular}
\end{table}

All image reductions and photometry were made using IRAF.\PZfm 
\PZfoot{IRAF is distributed by the National Optical Astronomy Observatory,
which is operated by AURA under cooperative agreement with the
National Science Foundation}

\medskip

The image of NGC 3147 obtained at C60 in the $R$ band is presented in
Fig.~1. SN 2008fv and the sites of three previous SNe are marked,
as well as the local standard stars.

\PZfig{12cm}{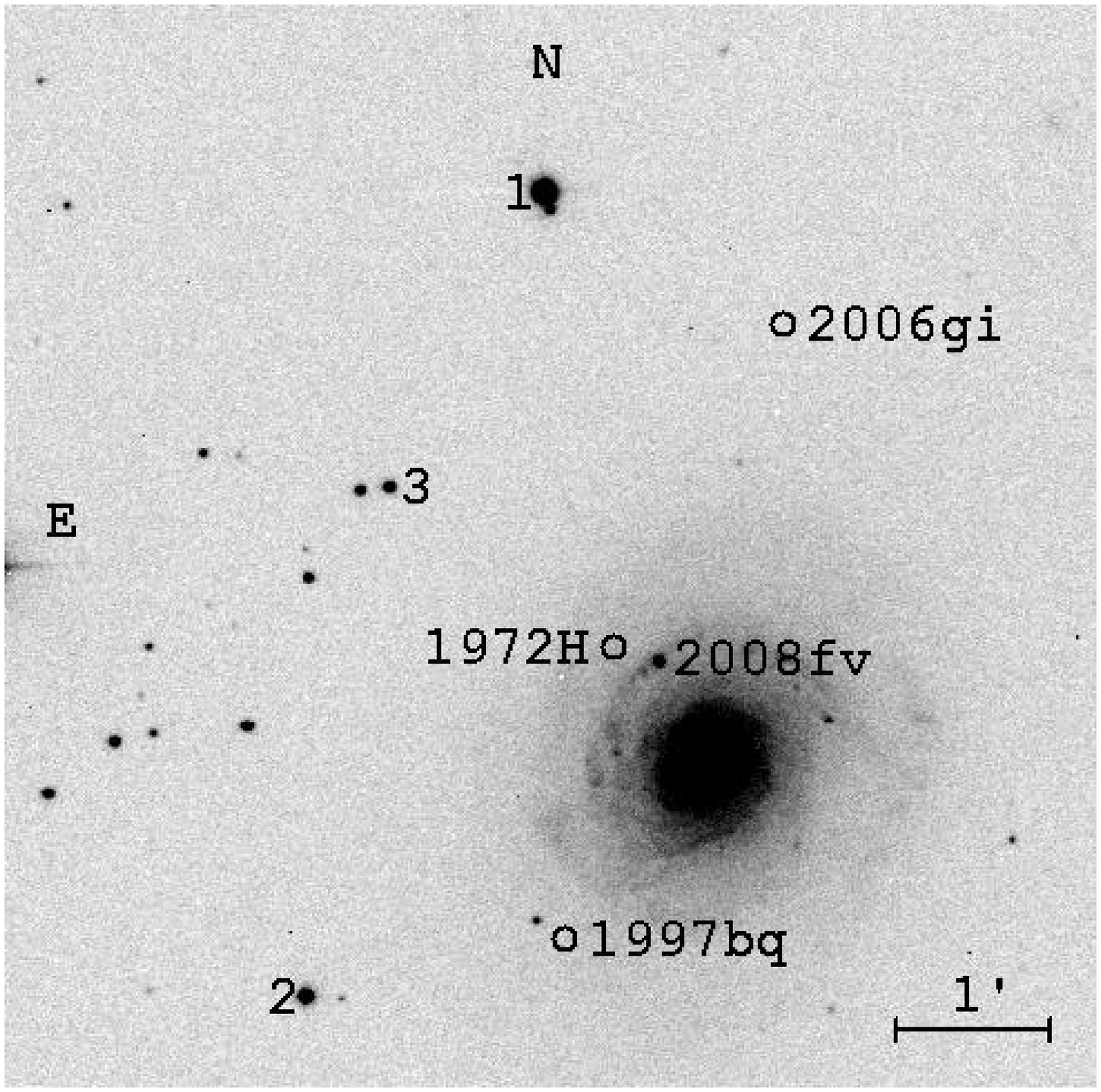}{SN 2008fv in NGC 3147, 
circles indicate 
the sites of previous SNe,
the local standard stars are marked}

The magnitudes of these stars are reported in Table~2, they were 
calibrated on 12 photometric nights during monitoring of SNe 2006gi
and 2008fv. Stars 1 and 3 were also used as standards for the photometry
of SN 2006gi (Elmhamdi et al., 2010). 

\begin{table}
\caption{Magnitudes of local standard stars}\vskip2mm
\begin{tabular}{ccccccccccc}
\hline
Star &$U$ & $\sigma_U$ & $B$ & $\sigma_B$ & $V$ & $\sigma_V$ & $R$ &
$\sigma_R$ & $I$ & $\sigma_I$ \\
\hline
1& 13.88& 0.05& 13.40& 0.01& 12.58& 0.01& 12.12& 0.01& 11.75& 0.01 \\ 
2&      &     & 14.91& 0.02& 14.15& 0.02& 13.71& 0.01& 13.37& 0.03 \\
3&      &     & 17.28& 0.05& 16.43& 0.04& 15.95& 0.03& 15.45& 0.04 \\
\hline
\end{tabular}
\end{table}

SN 2008fv is projected on the spiral arm, and the subtraction of
galaxy background is necessary for reliable photometry.
The template images were constructed from
frames obtained in November 2006 at C60 while monitoring SN 2006gi. 
After template subtraction the magnitudes of
the SN were derived by PSF fitting relative to a sequence of local standard
stars. 

The observations of SN 2008fv at TM35 in $VR$ filters are presented in 
Table~3, and the photometry in $UBVRI$ filters obtained with other telescopes
is reported in Table~4.

\begin{table}
\caption{Observations of SN 2008fv at TM35 }\vskip2mm
\begin{tabular}{cccccccccc}
\hline
JD 2454000+  & $V$ & $\sigma_V$ & $R$ & $\sigma_R$ &  
JD 2454000+  & $V$ & $\sigma_V$ & $R$ & $\sigma_R$ \\
\hline
742.86  &   14.89& 0.03 & 14.73& 0.02 & 773.93 &   15.36& 0.03 & 15.12& 0.06 \\ 
746.92  &   14.56& 0.02 & 14.44& 0.02 & 776.86 &   15.57& 0.03 & 15.16& 0.02 \\
747.90  &   14.52& 0.02 & 14.43& 0.01 & 777.85 &        &      & 15.18& 0.01 \\
748.91  &   14.47& 0.02 & 14.36& 0.01 & 778.98 &   15.65& 0.03 & 15.23& 0.01 \\
749.91  &   14.47& 0.02 & 14.37& 0.02 & 779.98 &   15.71& 0.03 &      &      \\
752.88  &   14.45& 0.02 & 14.39& 0.01 & 781.89 &   15.85& 0.03 & 15.41& 0.02 \\
755.94  &   14.52& 0.02 & 14.44& 0.03 & 786.93 &   16.18& 0.04 & 15.73& 0.04 \\
756.93  &   14.59& 0.05 & 14.56& 0.02 & 787.97 &   16.35& 0.03 & 15.78& 0.04 \\
759.82  &   14.72& 0.03 & 14.68& 0.02 & 791.98 &   16.40& 0.04 & 15.99& 0.06 \\
761.95  &   14.82& 0.03 & 14.81& 0.02 & 795.98 &   16.49& 0.04 & 16.08& 0.03 \\
762.92  &   14.91& 0.02 & 14.89& 0.02 & 799.02 &   16.62& 0.05 & 16.20& 0.02 \\
763.97  &   14.96& 0.02 & 14.97& 0.01 & 801.99 &   16.66& 0.05 & 16.33& 0.04 \\
766.93  &   15.12& 0.02 & 15.08& 0.02 & 803.00 &   16.48& 0.12 & 16.33& 0.03 \\
767.95  &   15.17& 0.03 & 15.10& 0.01 & 804.98 &        &      & 16.58& 0.11 \\
768.92  &   15.20& 0.02 & 15.10& 0.01 & 806.96 &   17.02& 0.10 & 16.46& 0.09 \\
771.93  &   15.31& 0.03 & 15.14& 0.01 &        &        &      &      &      \\
\hline
\end{tabular}
\end{table}

\begin{table}
\caption{Observations of SN 2008fv at other telescopes }\vskip2mm
\begin{tabular}{cccccccccccc}
\hline
JD 2454000+  & $U$ & $\sigma_U$ & $B$ & $\sigma_B$ &  $V$ & 
$\sigma_V$ & $R$ & $\sigma_R$ &  $I$ & $\sigma_I$ & Tel. \\
\hline
755.33 & 14.89& 0.07& 14.89& 0.03 & 14.37& 0.03 & 14.41& 0.02 & 14.65& 0.07 & M70 \\
759.66 & 14.97& 0.08& 15.03& 0.03 & 14.62& 0.03 & 14.60& 0.03 & 14.99& 0.05 & S50 \\
778.45 &      &     &      &      & 15.56& 0.03 & 15.20& 0.02 & 14.80& 0.05 & C50 \\
779.56 &      &     & 16.88& 0.04 & 15.65& 0.02 & 15.22& 0.02 & 14.68& 0.07 & C60 \\
780.49 &      &     & 17.00& 0.04 & 15.70& 0.02 & 15.31& 0.02 & 14.78& 0.03 & C60 \\
781.59 &      &     & 17.13& 0.03 & 15.79& 0.03 & 15.36& 0.02 & 14.80& 0.03 & C60 \\
782.41 &      &     & 17.05& 0.04 & 15.86& 0.03 & 15.44& 0.02 & 14.81& 0.04 & C60 \\
783.55 &      &     & 17.17& 0.05 & 15.91& 0.03 & 15.49& 0.02 & 14.91& 0.04 & C60 \\
784.49 &      &     & 17.25& 0.07 & 15.94& 0.03 & 15.56& 0.02 & 14.96& 0.05 & C60 \\
786.42 &      &     & 17.20& 0.06 & 16.04& 0.04 & 15.67& 0.02 & 15.09& 0.06 & C60 \\
795.36 &      &     & 17.67& 0.04 & 16.47& 0.03 & 16.14& 0.02 & 15.60& 0.05 & C60 \\
796.48 &      &     & 17.53& 0.03 & 16.51& 0.03 & 16.17& 0.03 & 15.66& 0.05 & C60 \\
799.41 &      &     &      &      &      &      & 16.28& 0.05 &      &      & S50 \\
845.22 &      &     &      &      &      &      & 17.87& 0.10 &      &      & M70 \\
866.21 &      &     &      &      &      &      & 18.08& 0.09 &      &      & M70 \\
868.20 &      &     &      &      &      &      & 18.28& 0.09 &      &      & M70 \\
894.36 &      &     &      &      & 18.98& 0.06 & 18.90& 0.07 &      &      & M70 \\
\hline
\end{tabular}
\end{table}
   
The light curves are shown in Fig.~2. The maximum phase in $V$ and $R$ bands
is well-sampled by the observations, and we can derive
dates and magnitudes of maximum light: $V_{max}=14.44$ on
JD 2454752, $R_{max}=14.37$ on the same date. The rate of early decline in
the $V$ band corresponds to $\Delta m_{15}(V)=0.65$.
From the list of type Ia SNe presented by Hicken et al. (2009)
we select three SNe with similar $\Delta m_{15}(V)$ and best-observed
light curves: 1992bc, 2006ax, 2007af (Hamuy et al., 1996;
Hicken et al., 2009). We fit the light curves of SN 2008fv
with those for three listed above SNe and find out that in the $V$ band the 
fits
for all three objects are equally good. In the $B$ band SN 1992bc provide
the best fit, while in the $R$ band SN 2006ax gives the best result.
The $I$-band light curve cannot be fitted well by any of these SNe,
but SN 2006ax provides slightly better match.   

\PZfig{12cm}{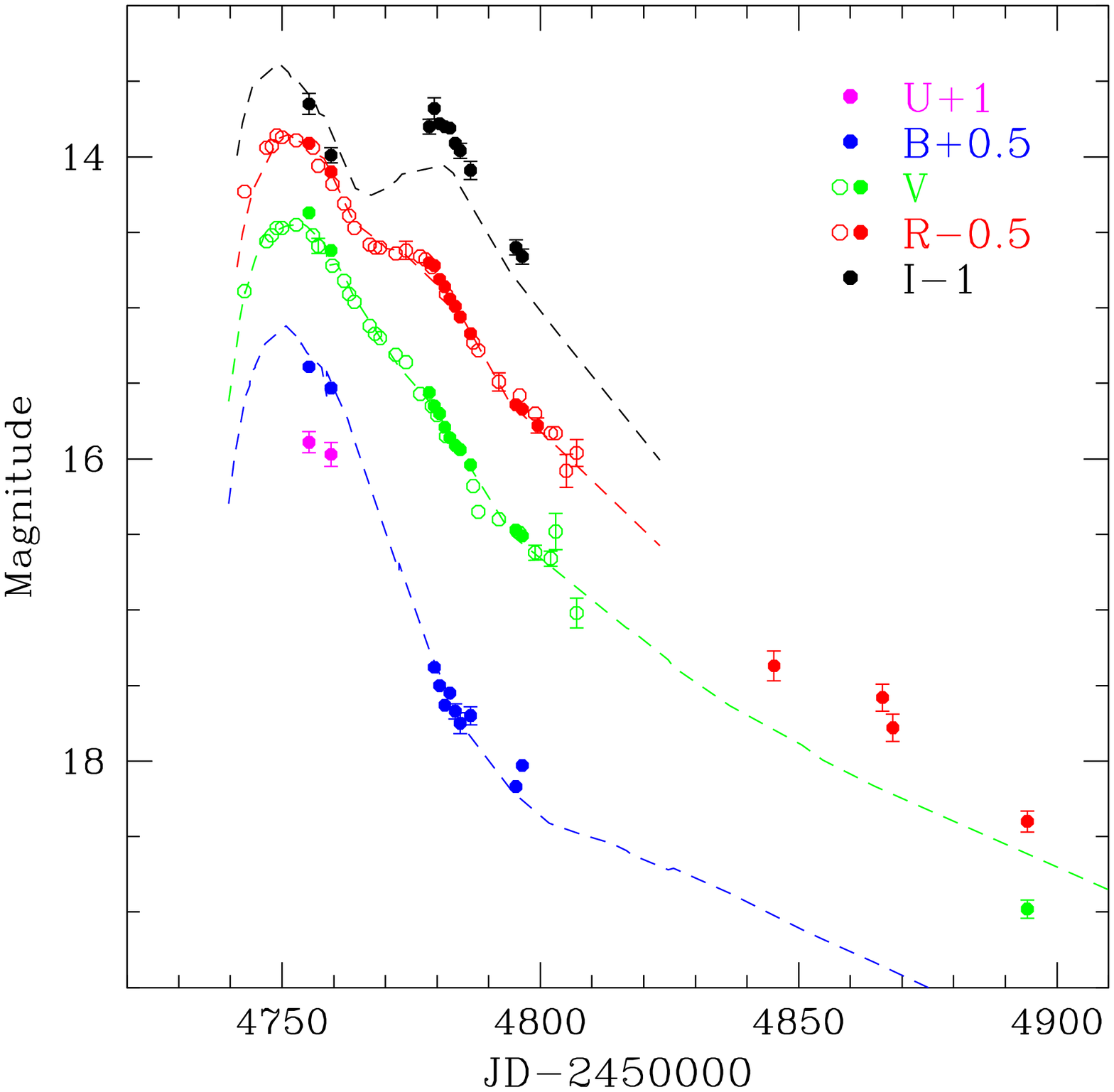}{The light curves of SN 2008fv.
Circles show data obtained at TM35, dots are for the observations
at other telescopes. 
The error bars are plotted only if they exceed the size of a symbol.
The dashed lines are the light curves of
SN 1992bc ($B$ and $V$ bands) and SN 2006ax ($R$ and $I$ bands)
}

After JD 2454820 the $V$ and $R$ light curves show the linear decline
with the rate 0.022 mag day$^{-1}$, which is quite normal for SNe Ia
at this stage.

\bigskip
\bigskip
{\PZsubtitle{SN 1972H}
SN 1972H was discovered by Goranskij (1972) on a plate exposed on 
1972 August 3.9 UT
at Crimean observatory of Sternberg Astronomical Institute.
Photographic photometry in the band close to $B$ was reported
by Goranskij (1972), and in $B,V$ filters by Barbon et al. (1973).
The magnitudes of the SN were visually estimated on the plates 
relative to the sequences of local 
comparison stars, which were calibrated photographically.
While compiling the 
data on the light curves of SNe, we recalibrated these standards using
photoelectric and photographic photometry and measured the
plates obtained by Goranskij (1972) with microphotometer (Tsvetkov, 1985).
Now we calibrate these stars again
on our CCD frames relative to our local standards 1 and 2.
The results confirm the conclusions by Tsvetkov (1985) that the
calibration of Barbon et al. (1973) in the $B$ band is quite
accurate, but in the $V$ their magnitudes are systematically fainter
by about 0.15 mag. But for some of the comparison stars 
the random errors of our photographic photometry are found to be quite
large. We combine old plate measurements with the new calibration
of local standards and obtain new magnitudes for SN 1972H, which are
reported in Table~5. We correct the magnitudes of Barbon et al. (1973) 
for the systematic errors of their comparison stars photometry. The
resulting light curves for SN 1972H are shown in Fig.~3. 

\PZfig{12cm}{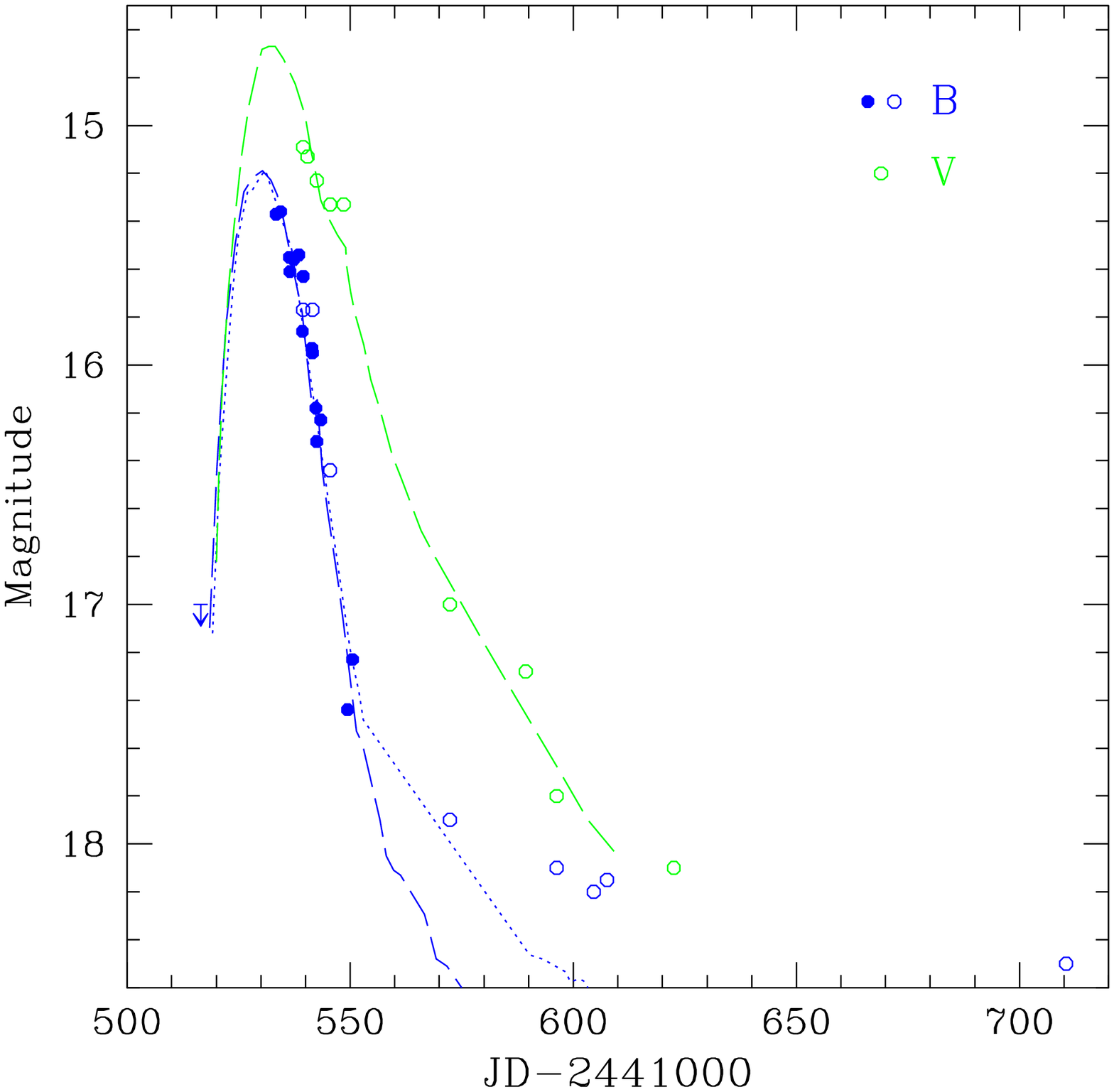}{$BV$ light curves of SN 1972H,
showing our photometry (dots) and that of Barbon et al. (1973)
(circles). Dashed lines are the $BV$ light curves of SN 1994D, 
dotted line is $B$-band light curve of SN 2007gi 
}

\begin{table}
\caption{Photographic observations of SN 1972H}\vskip2mm
\centering
\begin{tabular}{cccc}
\hline
JD 2441500+ & $B$  & JD 2441500+ & $B$  \\
\hline
16.5 & $>$17  &  39.5 &  15.63 \\  
33.4 & 15.37  &  41.4 &  15.93 \\
34.4 & 15.36  &  41.5 &  15.95 \\
36.4 & 15.55  &  42.3 &  16.18 \\
36.5 & 15.61  &  42.5 &  16.32 \\
37.3 & 15.56  &  43.4 &  16.23 \\
38.5 & 15.54  &  49.4 &  17.44 \\
39.3 & 15.86  &  50.5 &  17.23 \\
\hline
\end{tabular}
\end{table}

Unfortunately, no spectroscopic observations were reported for SN 1972H.
The classification as type I was proposed by Barbon et al. (1973) and was 
based on the shape of the light curve. We attempted to fit the data
for SN 1972H  
with the light curves of SNe Ia, IaPec(1991bg-class) and Ib/c, and
found that the best match is achieved for fast-declining SNe Ia,
such as 1994D and 2007gi (Richmond et al., 1995; Altavilla et al., 2004;
Zhang et al., 2010). Besides, there is a clear indication of a hump
on the $V$-light curve, which is never observed for type Ib/c or IaPec
SNe. So we may conclude that the probability of SN 1972H being normal
SN Ia is very high. The fitting of light curves of SNe 1994D and
2007gi indicate that SN 1972H reached maximum light around 
JD 2441530 with $B_{max}=15.2, V_{max}=14.7$.
The match for the $V$-band curve is good for all period covered by
observations, while in the $B$ band the tail of the light curve for 
SN 1972H is not fitted by the curve of SN 1994D, and only the start
of the tail is fitted by the curve of SN 2007gi. We may suppose that
the decline rate at the tail for SN 1972H is slower than usual, or that
there are large errors of the photographic magnitudes at late stage,
perhaps due to the galaxy background.

\bigskip
\bigskip
{\PZsubtitle{Comparison of light and color curves for three SNe}

The light curves of SNe 1972H, 1997bq (Jha et al., 2006) and 2008fv
in the $B,V$ bands are compared in Fig.~4.

\PZfig{12cm}{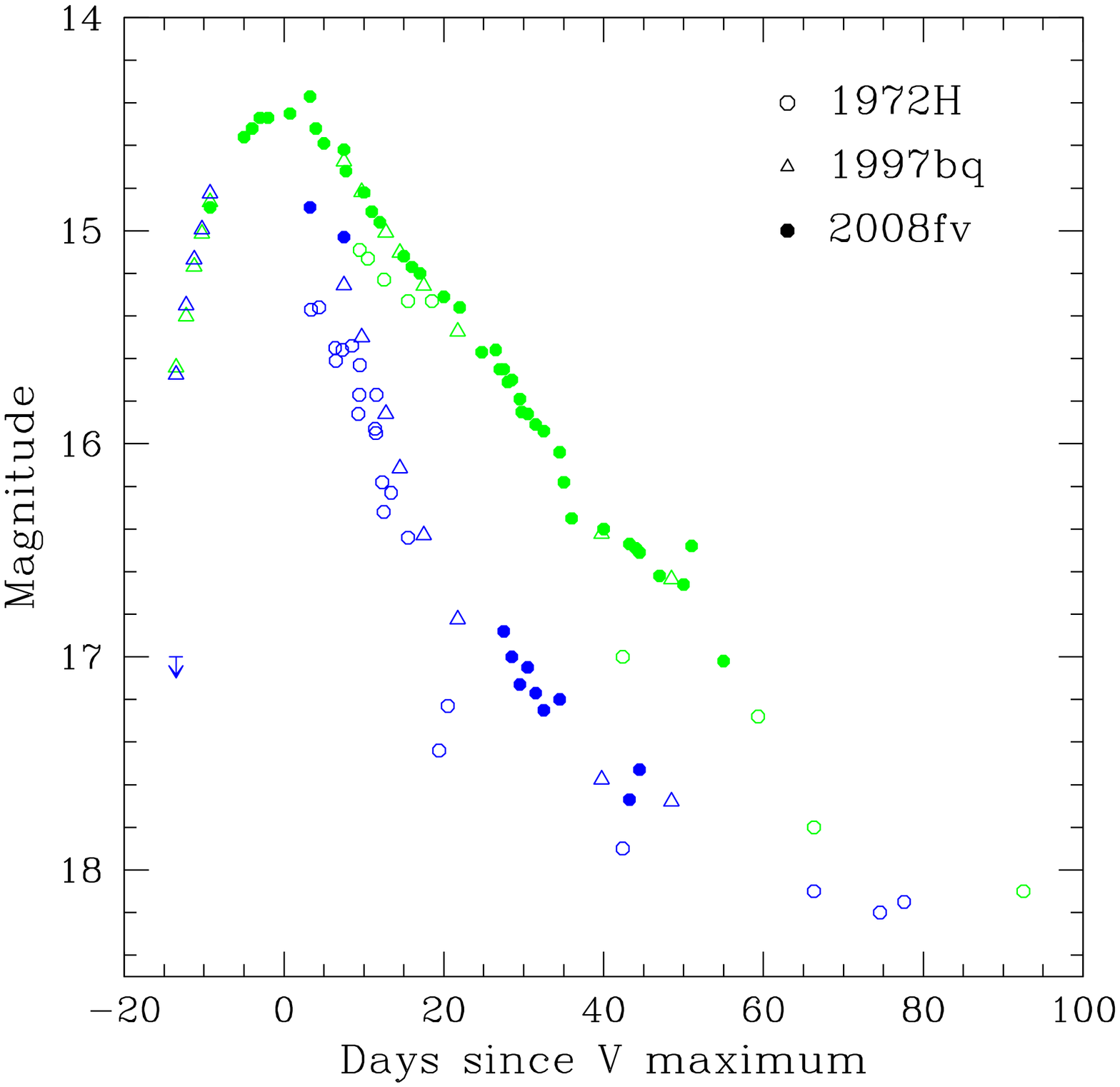}{The light curves of SNe 1972H, 1997bq and 
2008fv in $B$ (blue) and $V$ (green) bands.
}

The curves were shifted only in time to match the dates of maximum 
in the $V$ band. The striking similarity of $V$-light curves for 
SNe 1997bq and 2008fv is the most interesting result. The light curves
in $B$ band are also similar, but they are not so well-sampled, and 
the similarity is not so evident as for the $V$-curves. SN 1972H clearly shows
faster declining light curves with fainter maxima.
The light curves of SNe 1997bq and 2008fv in the $R$ and $I$ bands
are compared in Fig.~5. Again we see nearly perfect agreement between
these objects, there is only slight difference of the $R$-curves at 
phases 10-20 days past maximum. 

\PZfig{12cm}{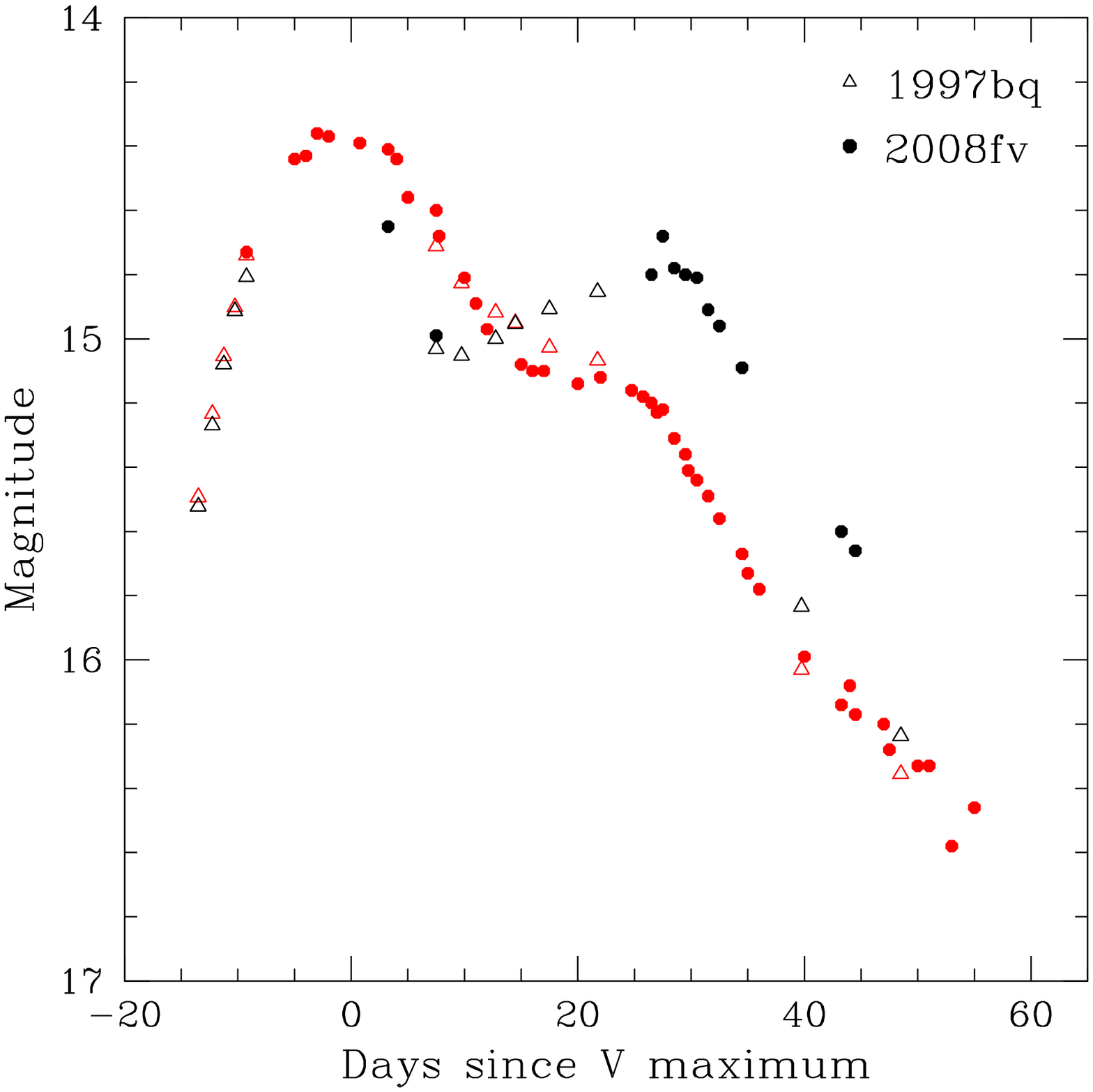}{The light curves of SNe 1997bq and
2008fv in $R$ (red) and $I$ (black) bands. 
}

The color curves are presented in Fig.~5. As expected, the color
curves of SNe 1997bq and 2008fv are nearly identical, the difference
can be noticed only on the $(V-R)$ curve in the period 10-20 days past maximum.
These SNe are clearly reddened, the comparison of $(B-V)$ color after
phase 30 days with the "Lira-Phillips relation"  
(Phillips et al., 1999)
allows to estimate
$E(B-V)=0.2$. The $(B-V)$ color of SN 1972H is significantly
redder than the curve for SN 1994D at the phase of early decline, but
later the color becomes even bluer than the "Lira-Phillips relation".
But we should note that colors derived from photographic photometry
may have large errors. So we cannot make any reliable estimate of
reddening for SN 1972H, and only suppose that $E(B-V)$ 
probably is in the range $0-0.2$ mag.

\PZfig{12cm}{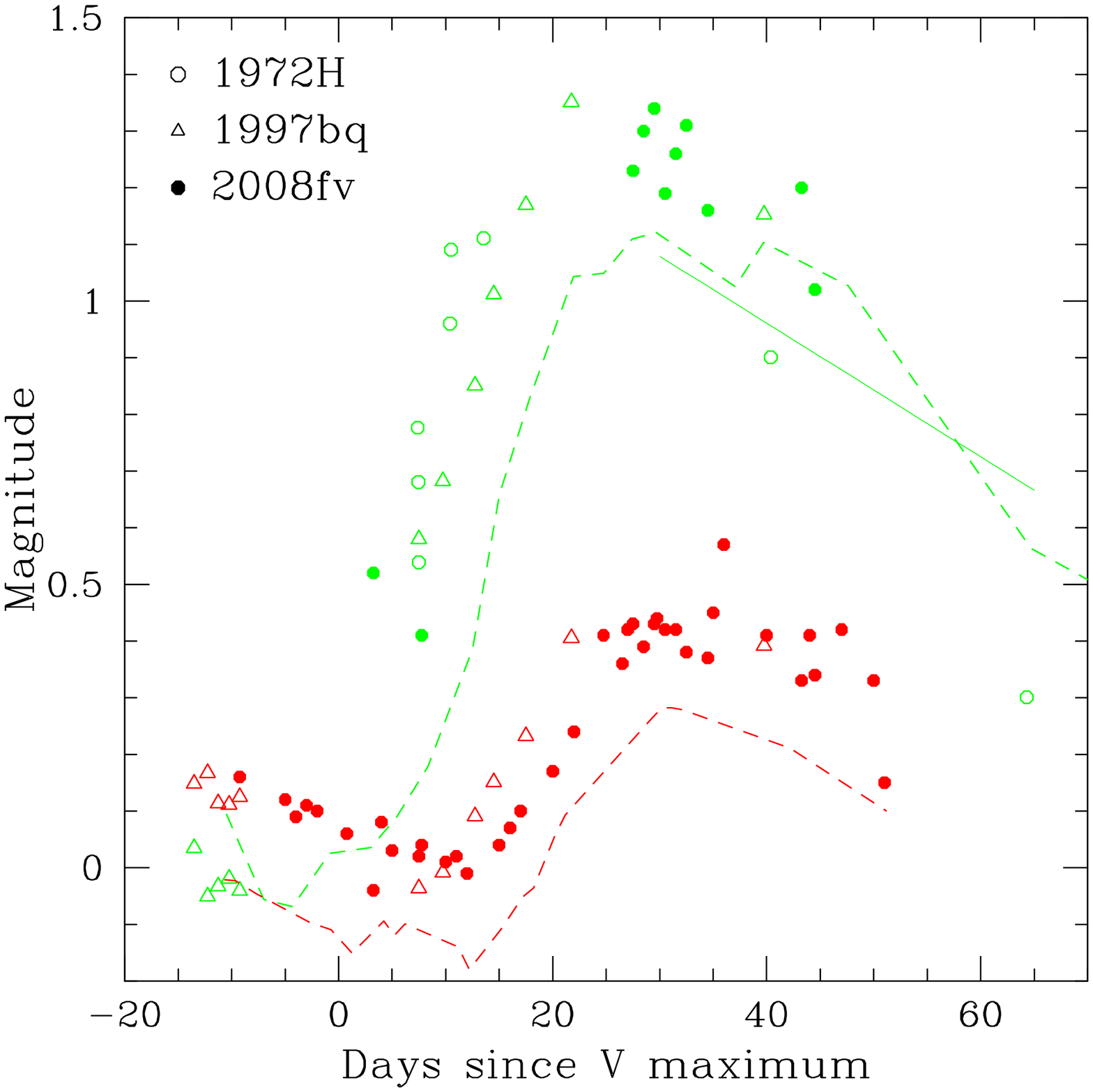}{The $(B-V)$ (green) and $(V-R)$ (red)
color curves for SNe 1972H, 1997bq and 2008fv. Green dashed line is
the $(B-V)$ color curve of SN 1994D, red dashed line is the 
$(V-R)$ color curve of SN 2006ax. The solid green line presents the
"Lira-Phillips relation".}

\medskip

Among the three SNe only for SN 2008fv the rate of decline in the $V$
band can be directly determined. Observations of this SN in $B$-band
started after maximum. SN 1972H was discovered after maximum, and we
have only an upper limit to constrain the rising branch of the light
curve. SN 1997bq have well-observed parts of the light curves before and
after maximum, but there is a gap around maximum light.

From the fitting of template light curves we can estimate that for 
SN 1972H $\Delta m_{15}(B)\approx 1.35,  \Delta m_{15}(V)\approx0.8$;
for SN 2008fv $\Delta m_{15}(B)\approx 0.85$.
Jha et al. (2006) estimate $\Delta m_{15}(B)=1.01$ for SN 1997bq, 
and we suppose that in the $V$-band its light curve is identical to the one
for SN 2008fv, with $\Delta m_{15}(V)=0.65$.  

If we assume distance modulus for NGC 3147 $\mu=33.2$ and the
galactic extinction $A_V^{gal}=0.077$, as in Elmhamdi et al. (2010),
the maximum absolute magnitudes for SNe 1997bq and 2008fv (uncorrected for 
extinction in
the parent galaxy) are $M_B=-18.7$, $M_V=-18.84$, and for 
SN 1972H $M_B=-18.1$, $M_V=-18.6$.  
If we compare these data with the plots of SNe Ia absolute magnitudes
versus $\Delta m_{15}(B)$ reported by Hicken et al. (2009), we find
that they are within the scatter of points presenting $M_B$ and $M_V$,
not corrected for the host-galaxy extinction, but are significantly
fainter than the mean values of $M_B^0$ and $M_V^0$.

After correction for the host galaxy extinction with $E(B-V)=0.2$
and $R_V=3.1$ we obtain for SNe 1997bq and 2008fv $M_B^0=-19.4, M_V^0=-19.38$,
which is very close to the ridge line of the relations between
$M_B^0, M_V^0$ and $\Delta m_{15}(B)$. For SN 1972H we can estimate the
host galaxy extinction, comparing its $M_B$ and $M_V$ with mean values of
$M_B^0$ and $M_V^0$ for $\Delta m_{15}(B)=1.35$. The comparison
yields $A_B\approx0.9$ and $A_V\approx0.5$, which correspond to the
color excess $E(B-V)\approx0.2$. Such color excess means that  
$(B-V)$ color for SN 1972H is significantly bluer than the "Lira-Phillips 
relation"
at the tail stage. But these colors  
may have large errors, and even
with smaller color excess (lower limit is about 0.1 mag) the absolute 
magnitudes
of SN 1972H are still within the range of dispersion on the relation 
between $M_B^0$, $M_V^0$ and $\Delta m_{15}(B)$. 
\medskip

Let us consider also the locations of SNe inside the host galaxy.
SN 2008fv is in the spiral arm at the distance $R=7.5$ kpc from
the center of the galaxy, 
while the radius of the galaxy is 24.5 kpc.
SN 1972H exploded in an interarm region only
3.2 kpc from the site of SN 2008fv, at $R=9.9$ kpc.
SN 1997bq was located outside the region of spiral structure, 
at $R=16.0$ kpc. And the only one core-collapse SN in NGC 3147, 
SN Ib 2006gi, exploded at $R=30.9$ kpc, well outside the boundaries
of the galaxy.  

\medskip
The comparison of photometric data for three type Ia discovered in
NGC 3147 reveals very high similarity between two of them:
SNe 1997bq and 2008fv. Not only 
photometric parameters, but the host-galaxy extinction is also
the same for the two objects. 
This is quite surprising, taking into account the different environments of
these objects in the host galaxy.
The maximum luminosity of 
SNe 1997bq and 2008fv fits very well to the relation of absolute 
magnitude versus the initial decline rate. The data for SN 1972H are of
lower quality and do not allow definite determination of photometric
parameters, but the most probable value of absolute magnitude is
also in agreement with the mentioned above relation. The results confirm
the usefulness of type Ia SNe for providing accurate distance
estimates required for measuring cosmological parameters.

\bigskip
{\bf Acknowledgements.} We are grateful to S.Yu.Shugarov and I.M.Volkov, who
made some of the observations.
The work of D.T. was partly supported by the 
Leading Scientific Schools Foundation
under grant NSh.433.2008.2.

\references
Altavilla, G., Fiorentino, G., Marconi, M., et al., 2004,
{\it MNRAS}, {\bf 349}, 1344

Barbon, R., Ciatti, F., Rosino, L., 1973, {\it Astron. Astrophys.}, {\bf 29},
57

Challis, P., 2008, {\it CBET}, No. 1522 

Elmhamdi, A., Tsvetkov, D., Danziger, I.J., Kordi, A., 2010, preprint,
arXiv:1001.0945

Goranskij, V.P., 1972, {\it Astron. Tsirk.}, No. 723, 1

Hamuy, M., Phillips, M.M., Suntzeff, N.B., et al., 1996,
{\it Astron. J.}, {\bf 112}, 2408

Hicken, M., Challis, P., Jha, S., et al., 2009, {\it Astrophys. J.},
{\bf 700}, 331 

Itagaki, K., 2006, {\it IAUC}, No. 8751

Jha, S., Kirshner, R.P., Challis, P., et al., 2006, {\it Astron. J.},
{\bf 131}, 527

Laurie, S., Challis, P., 1997, {\it IAUC}, No. 6616

Nakano, S., 2008, {\it CBET}, No. 1520 

Phillips, M.M., Lira, P.,
Suntzeff, N.B., Schommer, R.A., Hamuy, M., Maza, J., 1999,
{\it Astron. J.}, {\bf 118}, 1766 

Richmond, M.W., Treffers, R.R., Filippenko, A.V., et al., 1995,
{\it Astron. J.}, {\bf 109}, 2121 

Tsvetkov, D.Yu., 1985, {\it Variable Stars}, {\bf 22}, 191

Zhang, T., Wang, X., Li, W., et al., 2010, {\it PASP}, {\bf 122}, 1

\endreferences
\end{document}